\documentclass{PoS}

\title{BKT phase transitions in strongly coupled $3D$ $Z(N)$ LGT at finite 
temperature}

\ShortTitle{BKT phase transitions in strongly coupled $3D$ $Z(N)$ LGT at 
finite temperature}

\author{O. Borisenko\\
        BITP, National Academy of Sciences of Ukraine,
        03680 Kiev, Ukraine \\
        E-mail: \email{oleg@bitp.kiev.ua}}

\author{V. Chelnokov\\
        BITP, National Academy of Sciences of Ukraine,
        03680 Kiev, Ukraine\\
        E-mail: \email{vchelnokov@i.ua}}

\author{\speaker{G. Cortese}\\
        Instituto de F\'isica Te\'orica UAM/CSIC, Cantoblanco, E-28049 Madrid, 
        Spain, Departamento de F\'isica Te\'orica, Universidad de Zaragoza, 
        E-50009 Zaragoza, Spain\\
        E-mail: \email{cortese@unizar.es}}

\author{R. Fiore\\
        Dipartimento di Fisica, Universit\`a della Calabria, and INFN - Gruppo 
        Collegato di Cosenza I-87036 Rende, Italy\\
        E-mail: \email{fiore@cs.infn.it}}

\author{M. Gravina\\
        Department of Physics, University of Cyprus, P.O. Box 20357, Nicosia, 
        Cyprus\\
        E-mail: \email{gravina@ucy.ac.cy}}

\author{A. Papa\\
        Dipartimento di Fisica, Universit\`a della Calabria, and INFN - Gruppo 
        Collegato di Cosenza I-87036 Rende, Italy\\
        E-mail: \email{papa@cs.infn.it}}

\author{I. Surzhikov\\
        BITP, National Academy of Sciences of Ukraine,
        03680 Kiev, Ukraine \\
        E-mail: \email{i\_van\_go@inbox.ru}}

\abstract{We investigate, both analytically and numerically, the 
phase diagram of three-dimensional $Z(N)$ lattice gauge theories at finite 
temperature for $N>4$. These models, in the strong coupling limit, are 
equivalent to a generalized version of vector Potts models in two dimension, 
with Polyakov loops playing the role of $Z(N)$ spins. It is argued that the 
effective spin models have two phase transitions of infinite order (i.e. BKT). 
Using a cluster algorithm we confirm this conjecture, locate the position of 
the critical points and extract various critical indices.}

\FullConference{The 30 International Symposium on Lattice Field Theory - 
Lattice 2012,\\
		June 24-29, 2012\\
		Cairns, Australia}

\begin{document}

\section{Introduction and motivations}

After the discovery of the Berezinskii-Kosterlitz-Thouless (BKT) phase 
transition~\cite{Bere,Koste1,Koste2}, almost 40 years ago, this phenomenon 
still remains an interesting subject.

It is widely known that this kind of transition occurs in a variety of 
two-dimensional ($2D$) systems like, for instance, the most elaborated 
one represented by the $2D$ $XY$ model. 
However, there are several indications that the BKT phase transitions are also 
presents in some $3D$ lattice gauge models at finite temperature. Here we 
study $3D$ $Z(N)$ lattice gauge theories (LGT) at finite temperature in the 
strong coupling regime.

While the phase structure of $3D$ pure $Z(N)$ LGT for $N=2,3$ has been the 
subject of an intensive study, much less is known about the 
finite-temperature deconfinement transition when $N>4$. 
On the basis of the Svetitsky-Yaffe conjecture~\cite{Sve} that connect 
critical properties of $3D$ $Z(N)$ LGT with the corresponding properties of 
$2D$ spin models, we perform this study in order: 
\begin{itemize}
\item to clarify the order of the phase transitions that occur (if they
are of BKT-type) and then to check the prediction for the magnetic critical 
index $\eta$ and the compatibility with the $XY$ value for the index $\nu$;

\item to confirm the universality with $2D$ $Z(N)$ vector models and to 
provide checking-points of universality with $3D$ $SU(N)$ LGT in the strong 
coupling region.
  
\end{itemize}

\subsection{The Model}

We consider a $3D$ lattice $\Lambda = L^2\times N_t$ with spatial (temporal) 
extension $L$ ($N_{t}$); $\vec{x}=(x_0,x_1,x_2)$ where $x_0\in [0,N_t-1]$
and $x_1,x_2\in [0,L-1]$ represent the sites of the lattice
and $e_n$ the unit vector in the $n$-th direction. 
We denote $p_t$ ($p_s$) temporal (spatial) plaquettes, $l_t$ ($l_s$) temporal 
(spatial) links and periodic boundary conditions on gauge fields are imposed 
in all directions. The conventional plaquette angles $s(p)$ is
\begin{equation}
s(p) \ = \ s_n(x) + s_m(x+e_n) - s_n(x+e_m) - s_m(x) \ .
\label{plaqangle}
\end{equation}
The partition function can be expressed as
\begin{equation}
Z(\Lambda ;\beta_t,\beta_s;N) \ = \  \prod_{l\in \Lambda}
\left ( \frac{1}{N} \sum_{s(l)=0}^{N-1} \right ) \ \prod_{p_s} Q(s(p_s)) \
\prod_{p_t} Q(s(p_t)) \, ,
\label{PTdef}
\end{equation}
where the most general $Z(N)$-invariant Boltzmann weight with $N-1$ different 
couplings is
\begin{equation}
Q(s) \ = \
\exp \left [ \sum_{k=1}^{N-1} \beta_p(k) \cos\frac{2\pi k}{N}s \right ] \ .
\label{Qpgen}
\end{equation}
The Wilson action corresponds to the choice $\beta_p(1)=\beta_p$, 
$\beta_p(k)=0, k=2,...,N-1$.

To study the $3D$ $Z(N)$ LGT in the strong coupling limit ($\beta_s = 0$) 
one can map the gauge model to a generalized $2D$ spin $Z(N)$ model with the 
action 
\begin{equation}
\label{modaction}
S \ =\ \sum_{x}\ \sum_{n=1}^2 \sum_{k = 1}^{N-1} \ \beta_k \  
\cos \left( \frac{2 \pi k}{N} \left(s(x) - s(x+e_n) \right) \right) \ .
\end{equation}
The effective coupling constants $\beta_k$ are derived from the coupling 
constant $\beta_t \equiv \beta$ of the $Z(N)$ LGT, using the following 
equation (the Wilson action is used for the gauge model):
\begin{equation}
\beta_k \ =\ \frac{1}{N} \sum_{p = 0}^{N - 1} \ln(Q_p) \cos \left(\frac{2 \pi 
p k}{N} \right) \ ,
\label{couplings}
\end{equation}
where
\begin{eqnarray}
\label{Qk}
Q_k&\ =\ &\sum_{p = 0}^{N - 1} \ \left(\frac{B_p}{B_0}\right)^{N_t}  \ 
\cos \left(\frac{2 \pi p k}{N} \right) \ ,   \\
B_k&\ =\ &\sum_{p = 0}^{N - 1} \exp \left[ \beta \cos \left(\frac{2 \pi p}{N} 
\right) \right ] 
\cos \left(\frac{2 \pi p k}{N} \right) \ .
\label{couplings_coeff}
\end{eqnarray}

In the strongly coupled $3D$ $Z(N)$ LGT one expects a scenario with three 
phases. Therefore two phase transitions must separate these three phases
when $N>4$ (BKT for $N\geqslant5$):
\begin{itemize}
\item transition from high-temperature to massless phase with $\eta=1/4$;
\item transition from massless phase to low-temperature phase with the 
prediction $\eta=4/N^{2}$.
\end{itemize}




\subsection{Observables}
Below are listed all the observables used in this work:
\begin{itemize}
\item the absolute value of the complex magnetization:
\begin{equation}
M_L \ =\  \sum_{x \in \Lambda} \exp \left( \frac{2 \pi i}{N} s(x) \right)
=|M_{L}|e^{i\psi}\,;
\nonumber
\end{equation}

\item the real part of the "rotated" magnetization $M_{R}=|M_{L}|\cos(N\psi)$
and the normalized rotated magnetization $m_\psi = \cos(N \psi)$\,;
\item the quantity called ``population'':
\begin{equation}
S_{L}=\frac{N}{N-1}\left( \frac{\max_{i=0,N-1}(n_{i})}{L^{2}}-\frac{1}{N} 
\right)\,,
\nonumber
\end{equation}
where $n_i$ is number of $s(x)$ equal to $i$;

\item the related susceptibilities $\chi_{L}^{(M)}$, $\chi_{L}^{(M_{R})}$, 
$\chi_{L}^{(S)}$ of the real part of the complex magnetization, of the 
rotated magnetization and of the population $S_{L}$, respectively;
\end{itemize}

\begin{itemize}
\item the reduced fourth-order Binder cumulant $U_{L}^{(M)}$ defined as 
\begin{equation}
U_{L}^{(M)}=1 - \frac{\langle|M_{L}|^{4}\rangle}{3\langle|M_{L}|^{2}
\rangle^{2}}\,;
\end{equation}
\nonumber
\item the cumulant $B_{4}^{(M_{R})}$ defined as
\begin{equation}
B_{4}^{(M_{R})}=\frac{\langle |M_{R} - \langle M_{R} \rangle|^{4} \rangle}
{\langle |M_{R} - \langle M_{R} \rangle|^{2} \rangle^{2}}\,;
\nonumber
\end{equation}
\item the helicity modulus~\cite{Koste3}
\begin{equation}
\Upsilon=\left<e\right> - L^{2}\beta\left<s^{2}\right>\,,
\nonumber
\end{equation}
where $e\equiv\frac{1}{L^{2}} \sum_{<ij>_{x}} \cos\left(\theta_{i}-\theta_{j}
\right)$\, and $s\equiv\frac{1}{L^{2}} \sum_{<ij>_{x}} \sin(\theta_{i}
-\theta_{j})$ with\,\, $\theta_i \equiv \frac{2\pi}{N} s(i)$.
\end{itemize}

We have simulated models with $N$=5,7,9,13 on lattices ranging from $L=128$
to $L=2048$.

\section{Results} 
\subsection{Determination of the critical temperatures}

In order to extract the critical indices we need to locate the critical 
temperatures. Below we list the methods used to do that 
for the first critical coupling $\beta_{\rm c}^{(1)}$: 

(a) we locate the positions of the $\beta_{\rm pc}^{(1)}(L)$ from the peak of 
the susceptibility $\chi_L^{(M)}$ of $|M_L|$ on various lattice sizes
and we find $\beta_{\rm c}^{(1)}$ by a fit with 
the following scaling function (with $\nu$ equal to 1/2):
\begin{equation}
\beta^{(1)}_{\rm pc}=\beta^{(1)}_{\rm c}
+\frac{A}{(\ln L + B)^{\frac{1}{\nu}}}\quad ;
\label{b_pc}
\nonumber
\end{equation}

(b) we estimate the crossing point of the
Binder cumulant $U_L^{(M)}$ versus $\beta$ on different lattices or, 
alternatively, we search for the value of 
$\beta_{\rm c}^{(1)}$ which optimizes the overlap of these curves when they 
are plotted against $(\beta-\beta_{\rm c}^{(1)})(\ln L)^{1/\nu}$ ($\nu$=1/2); 

(c) we consider the helicity modulus $\Upsilon$ near the phase transition and 
define $\beta_{\rm pc}^{(1)}(L)$ as the value of $\beta$ such that
$\eta(\beta) \equiv 1/(2 \pi \beta \Upsilon)=1/4$
on the various lattices; we then find $\beta_{\rm c}^{(1)}$ through the 
function 
\begin{equation}
\beta^{(1)}_{\rm pc}=\beta^{(1)}_{\rm c} + \frac{A}{\ln L + B}\;,
\label{helicity_scaling}
\end{equation}
valid under the assumption that the phase transition belongs to the $XY$ 
universality class.

To determine $\beta_{\rm c}^{(2)}$ we use:

(d) the same as the method (a) using instead the susceptibility $\chi_L^{(S)}$ 
of the population $S_L$;

(e) the same as the method (b) using instead simultaneously the Binder 
cumulant $B_4^{(M_R)}$ and the order parameter $m_\psi$. 

\begin{figure}
\begin{center}
\includegraphics[scale=0.50]{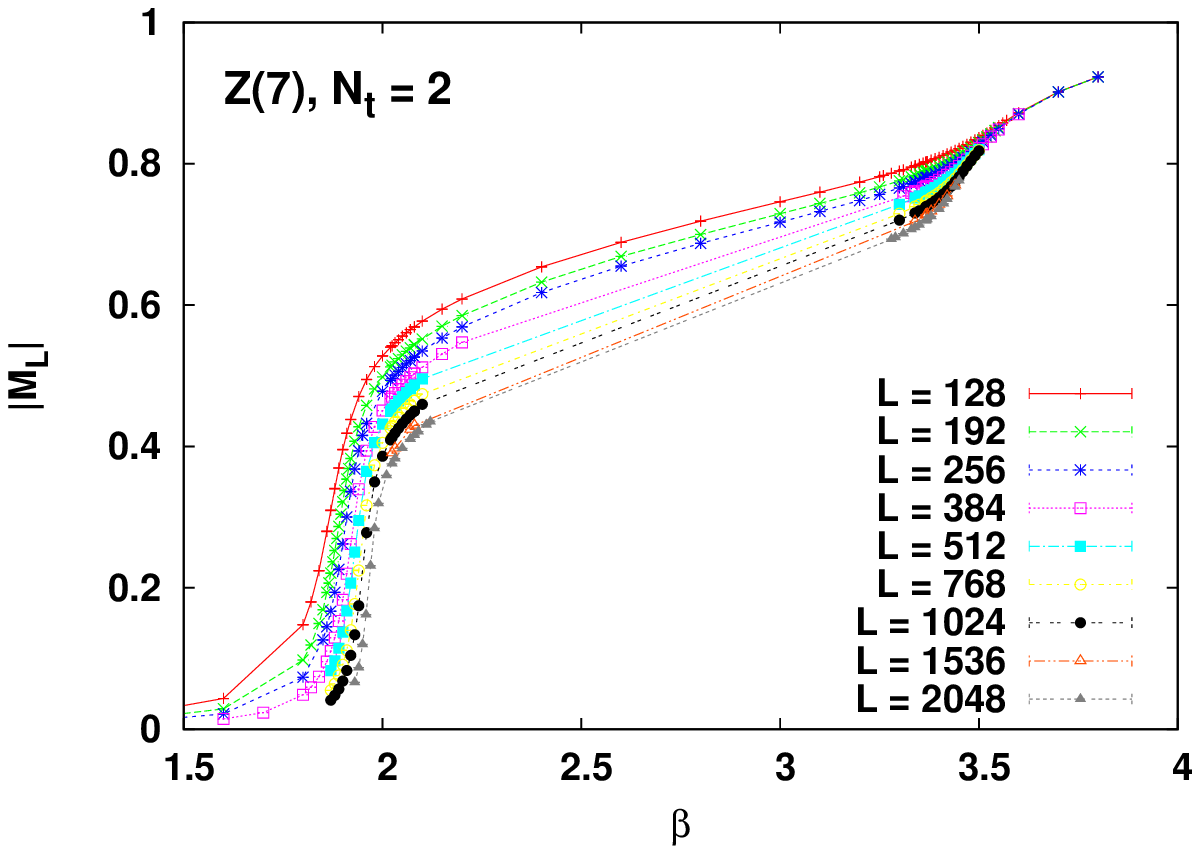}
\includegraphics[scale=0.50]{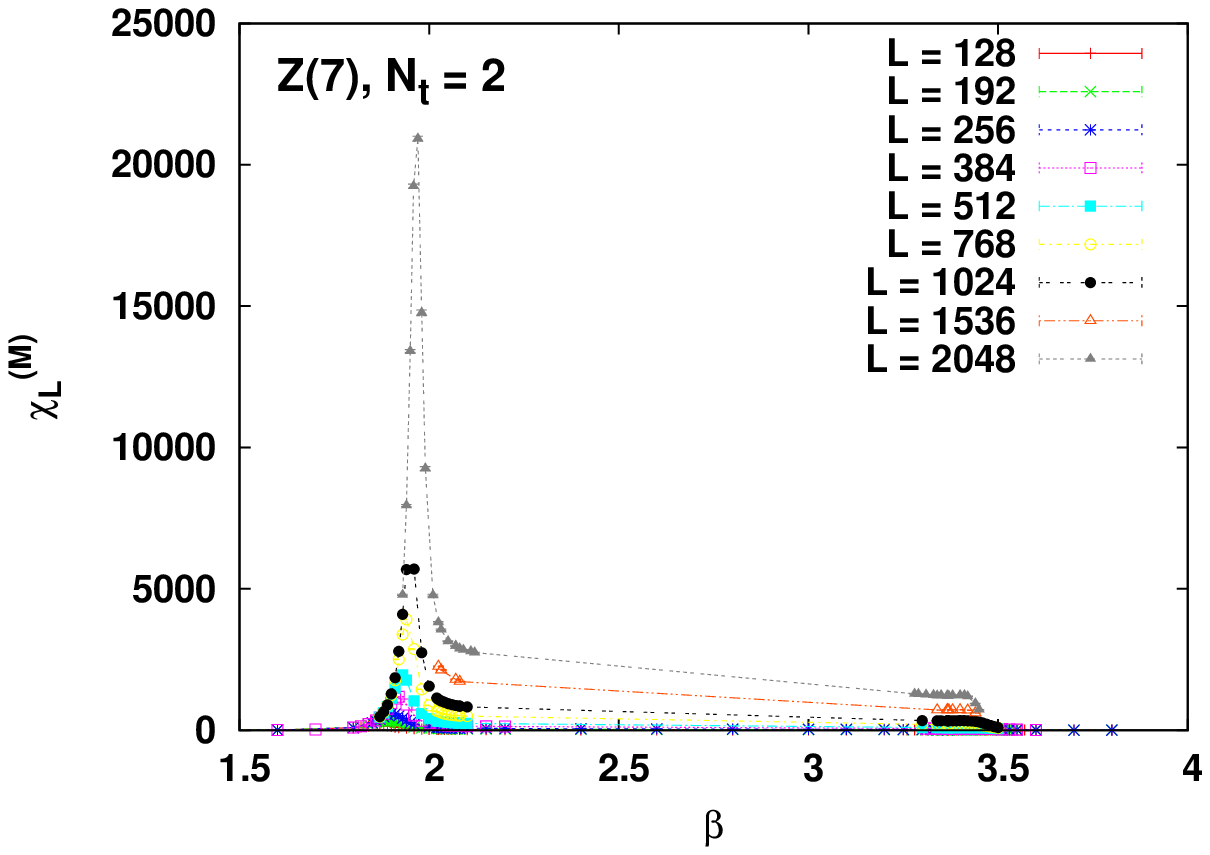}

\includegraphics[scale=0.50]{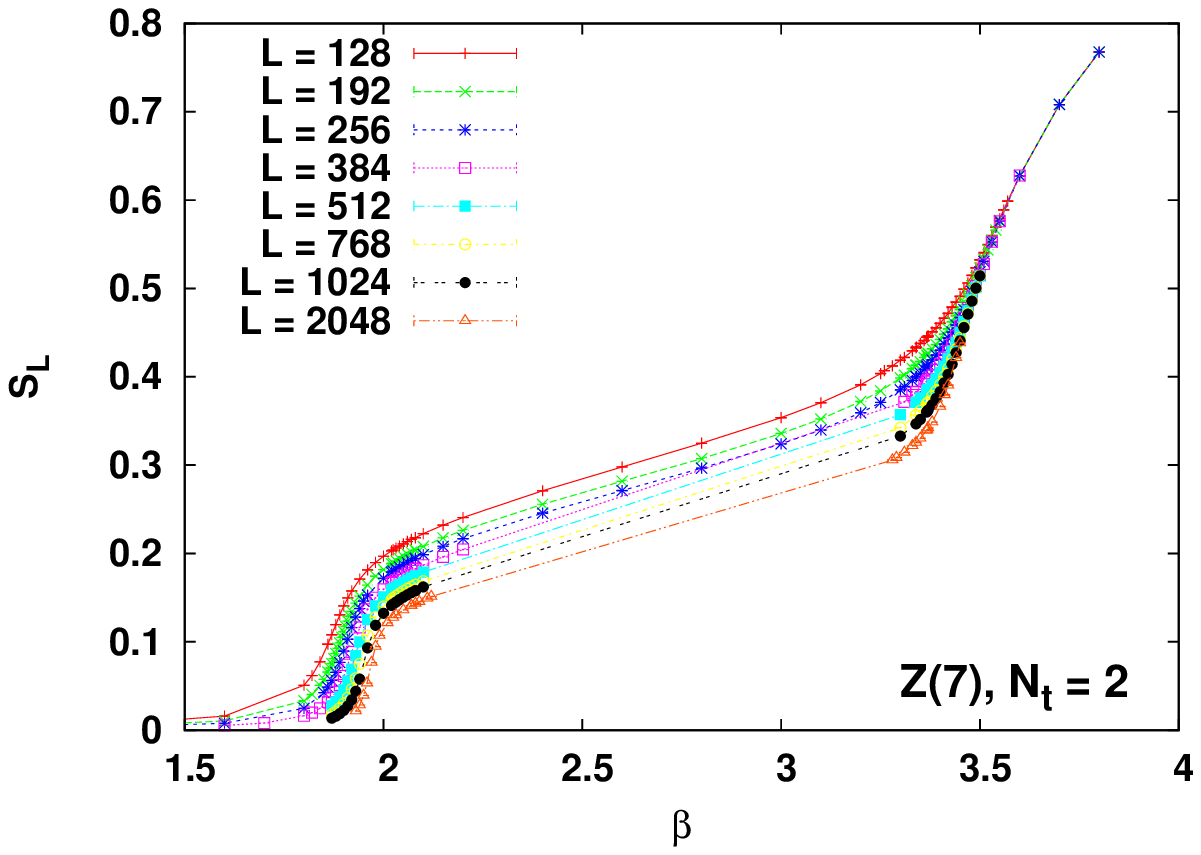}
\includegraphics[scale=0.50]{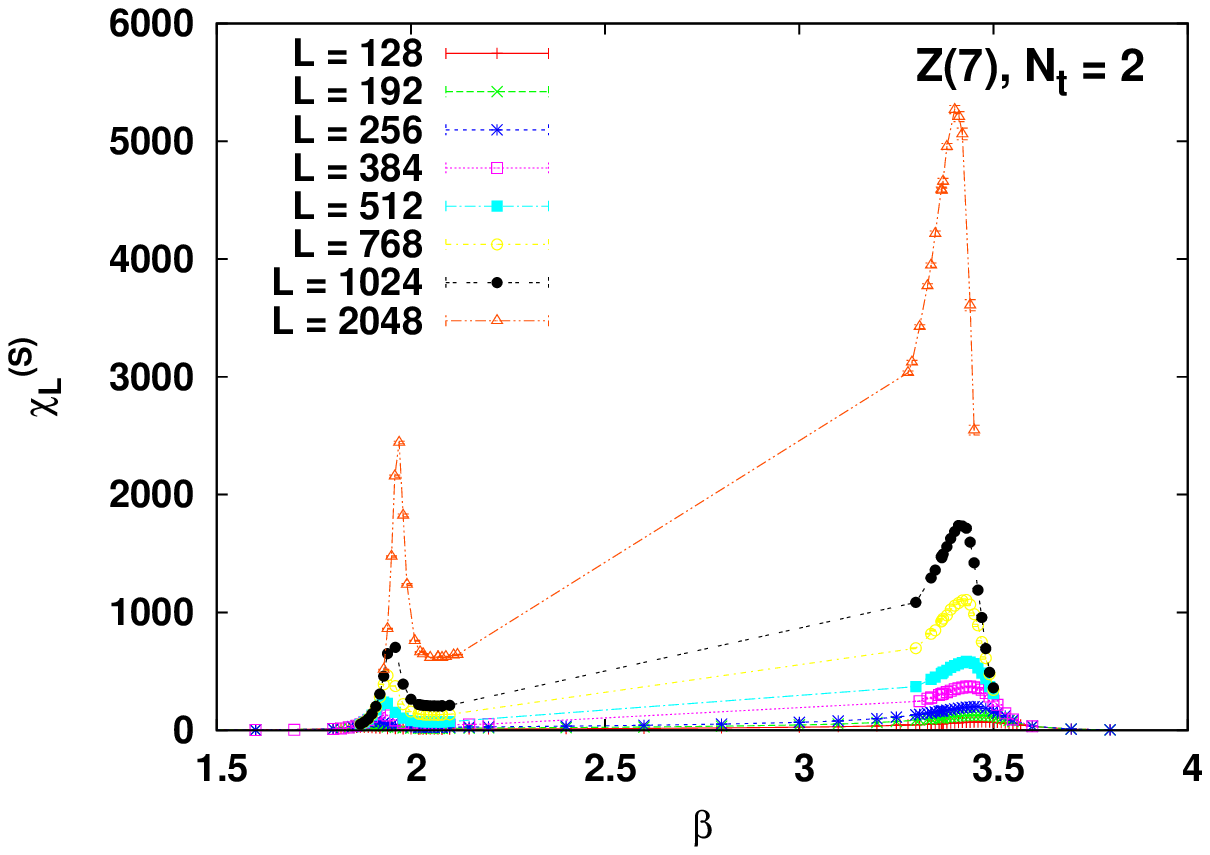}
\end{center}
\caption{The two plots in the first line illustrate the method (a),
while those on the second line illustrate the method (d).}
\label{figura1}
\end{figure}

\begin{figure}
\begin{center}
\includegraphics[scale=0.38]{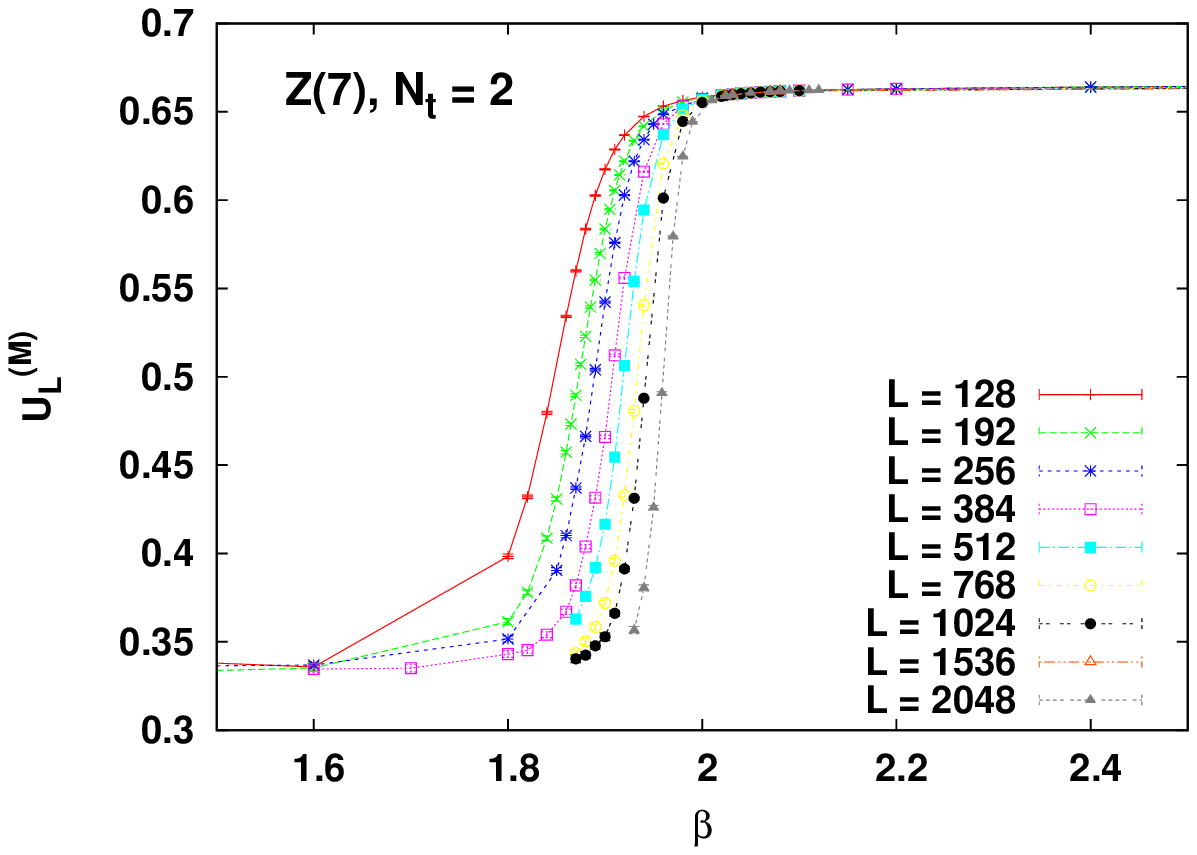}
\includegraphics[scale=0.38]{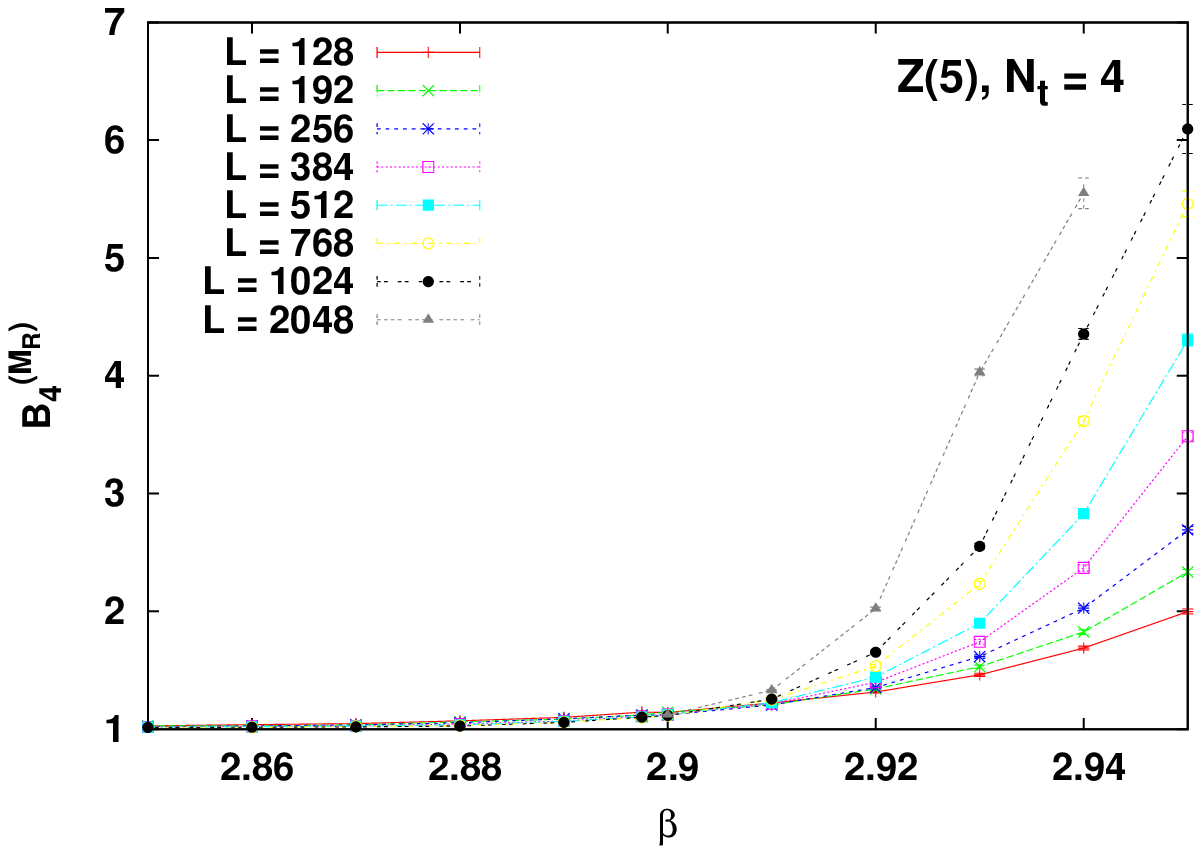}
\includegraphics[scale=0.38]{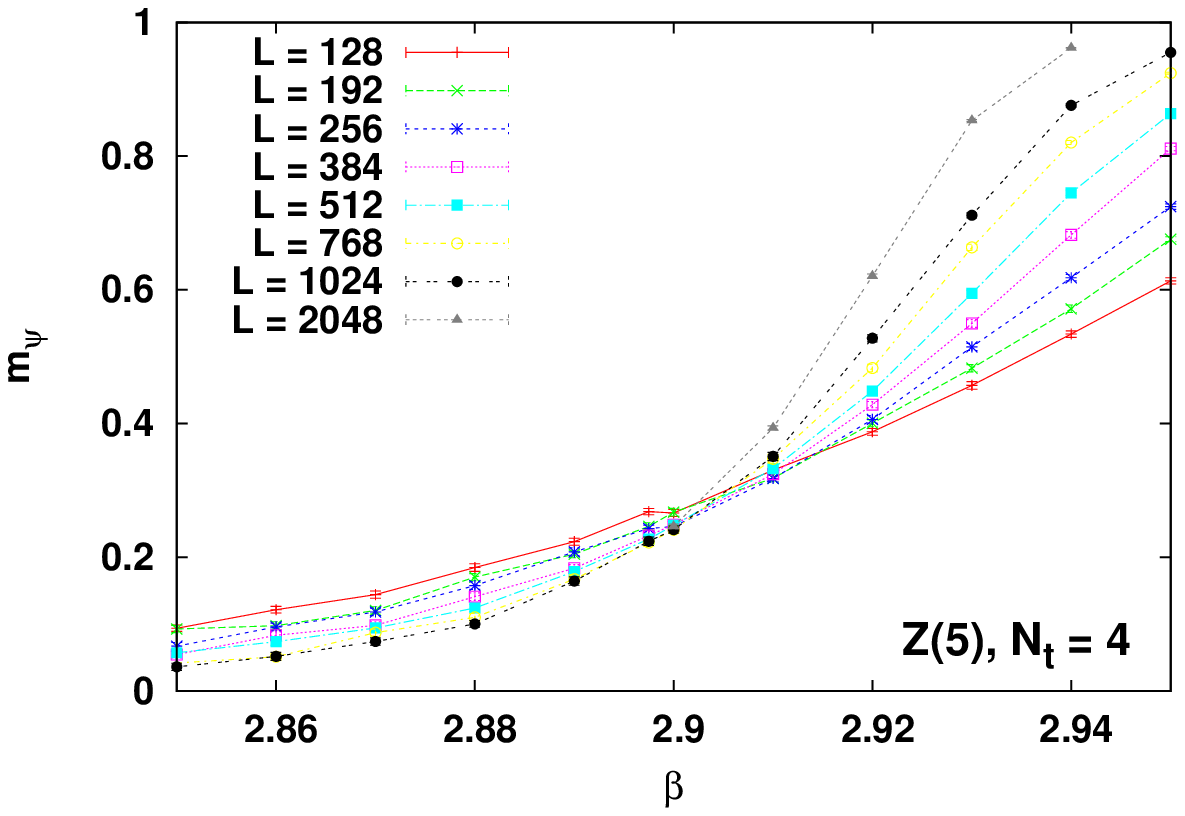}

\includegraphics[scale=0.38]{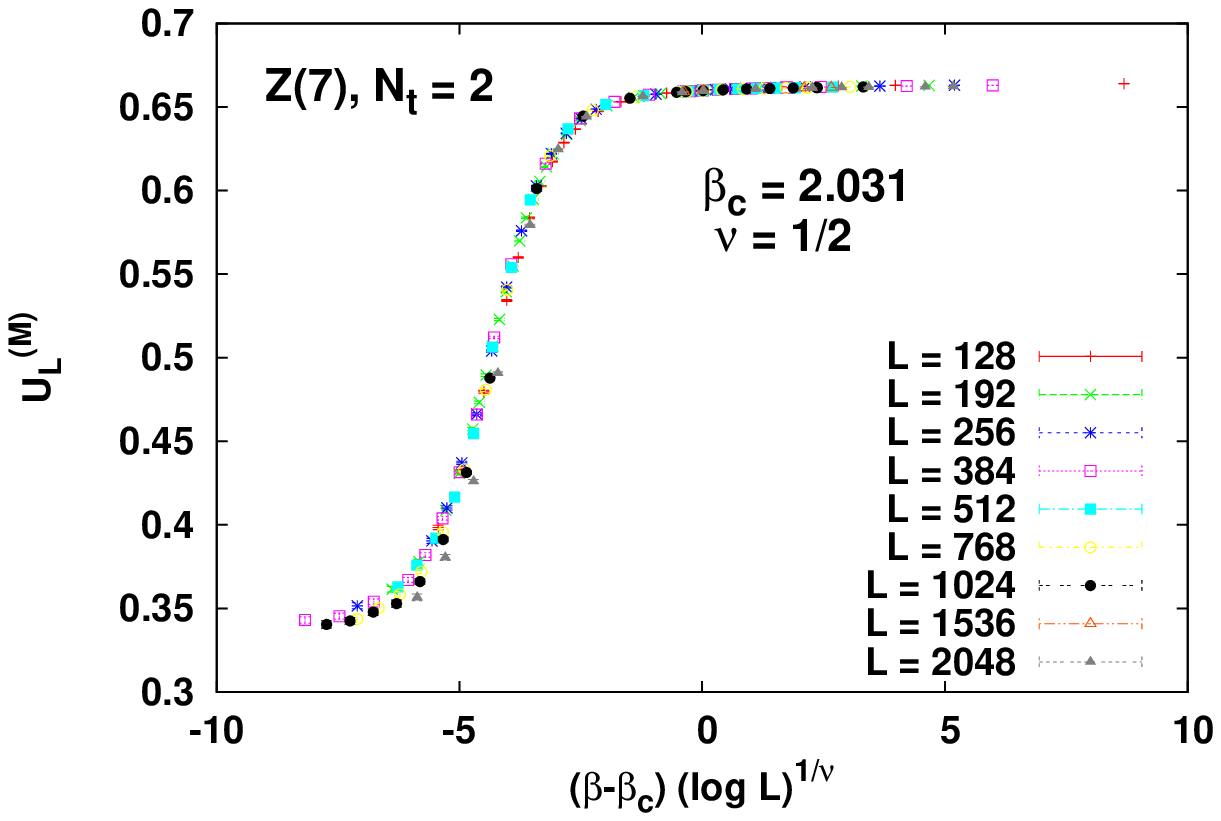}
\includegraphics[scale=0.38]{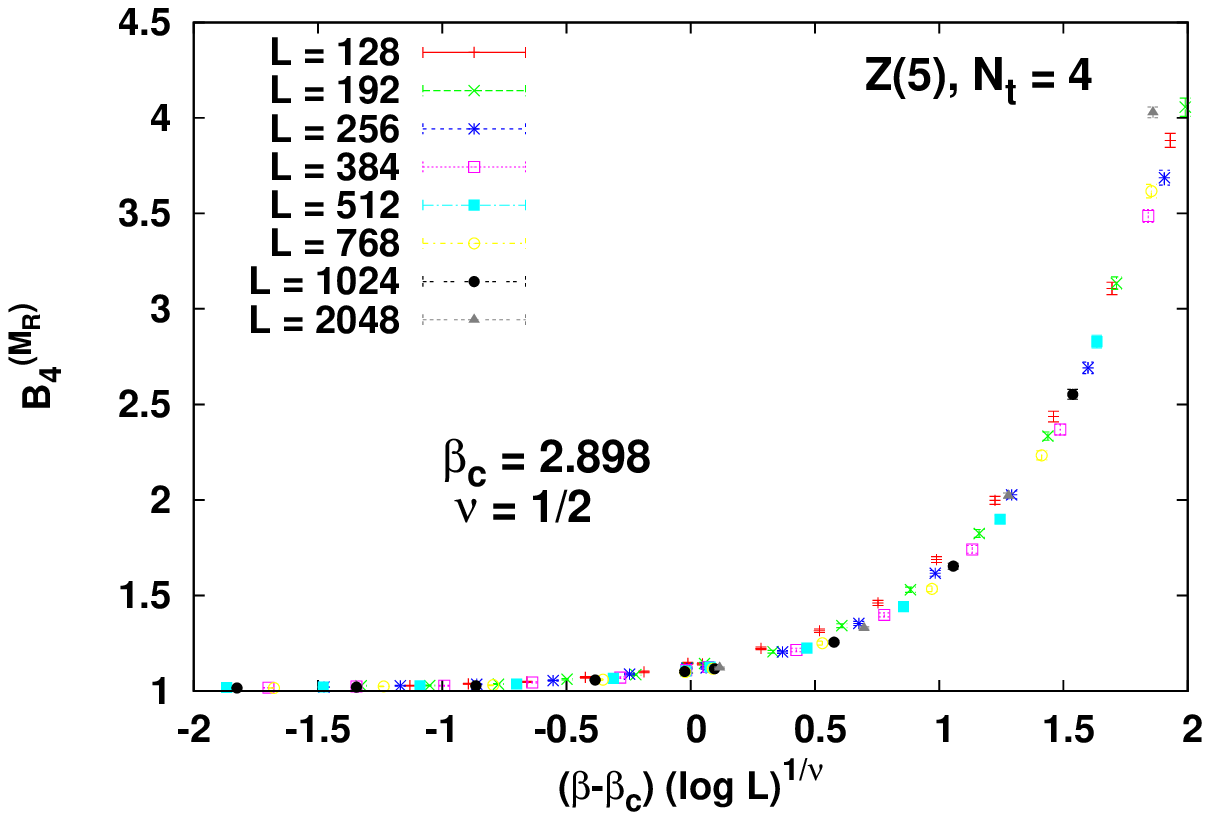}
\includegraphics[scale=0.38]{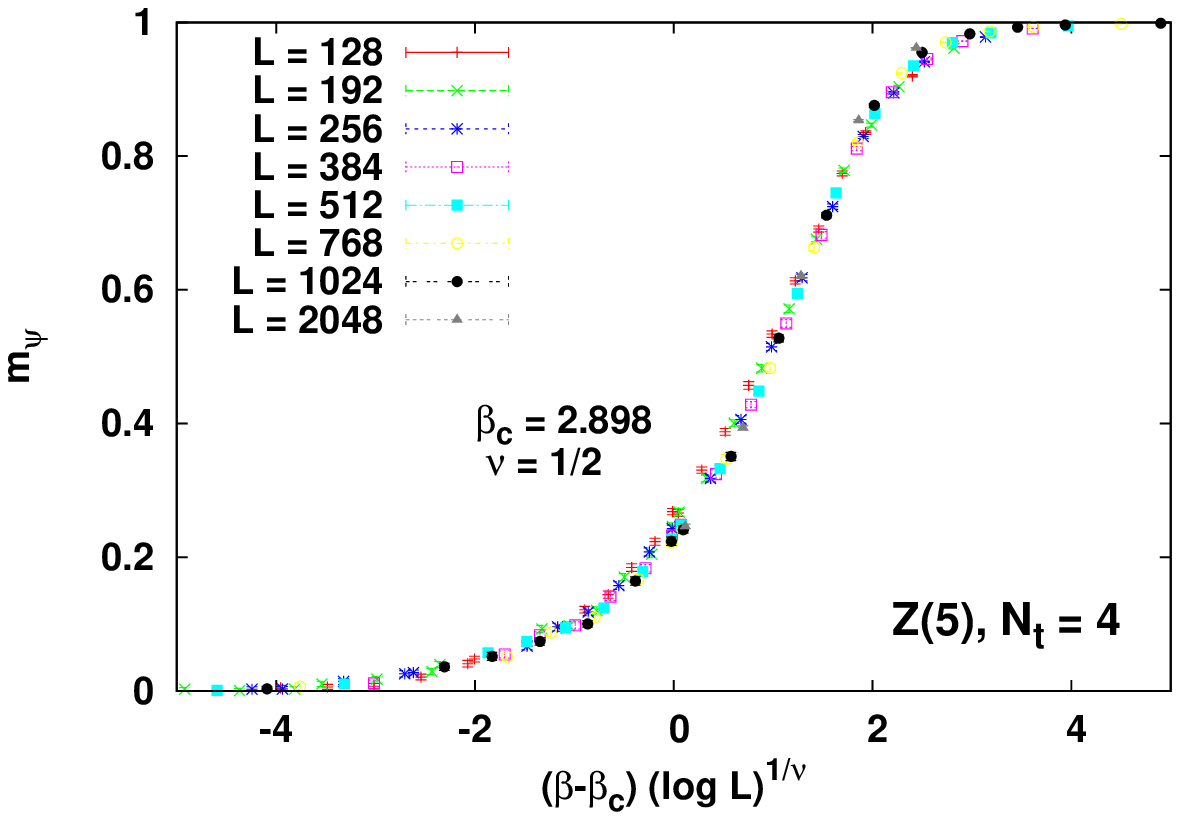}

\end{center}
\caption{The two plots in the first column illustrate the method (b),
while the remaining are illustrations of the method (e).}
\label{figura2}
\end{figure}

\begin{figure}
\begin{center}
\includegraphics[scale=0.50]{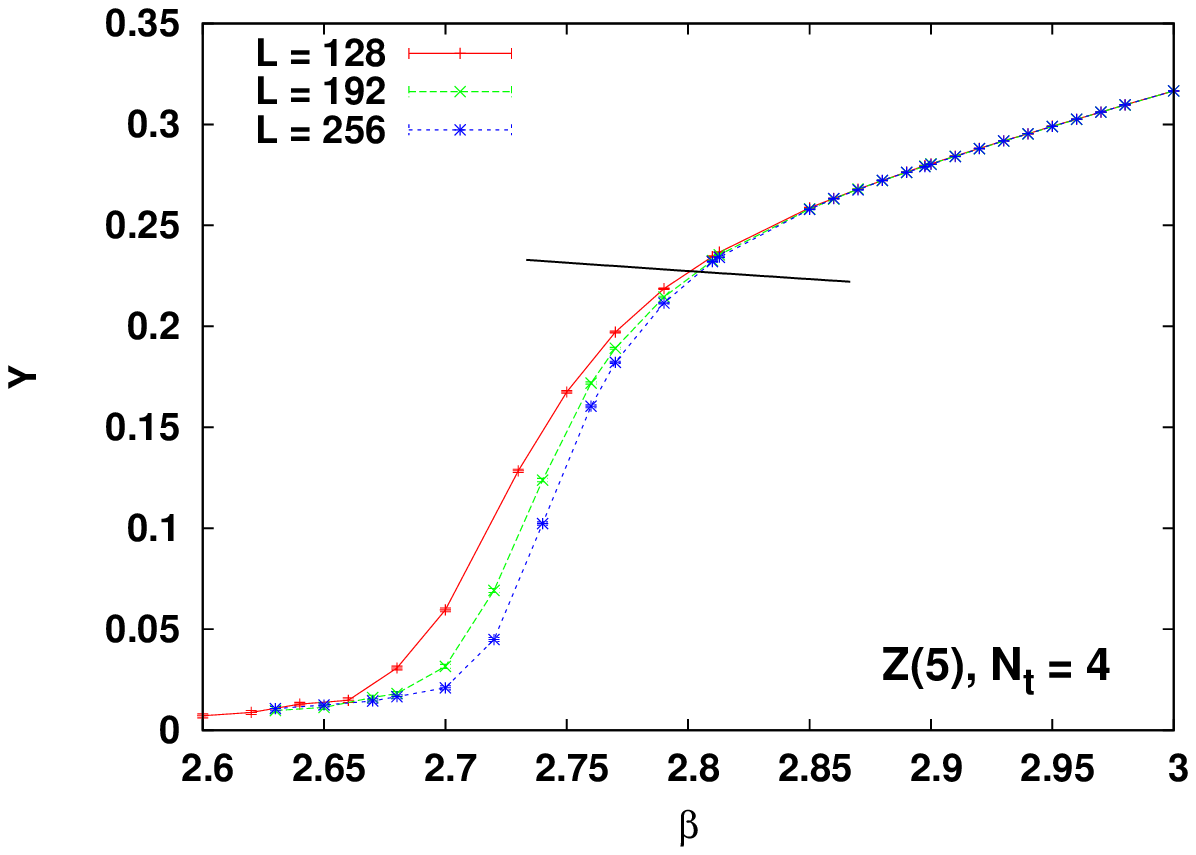}
\includegraphics[scale=0.50]{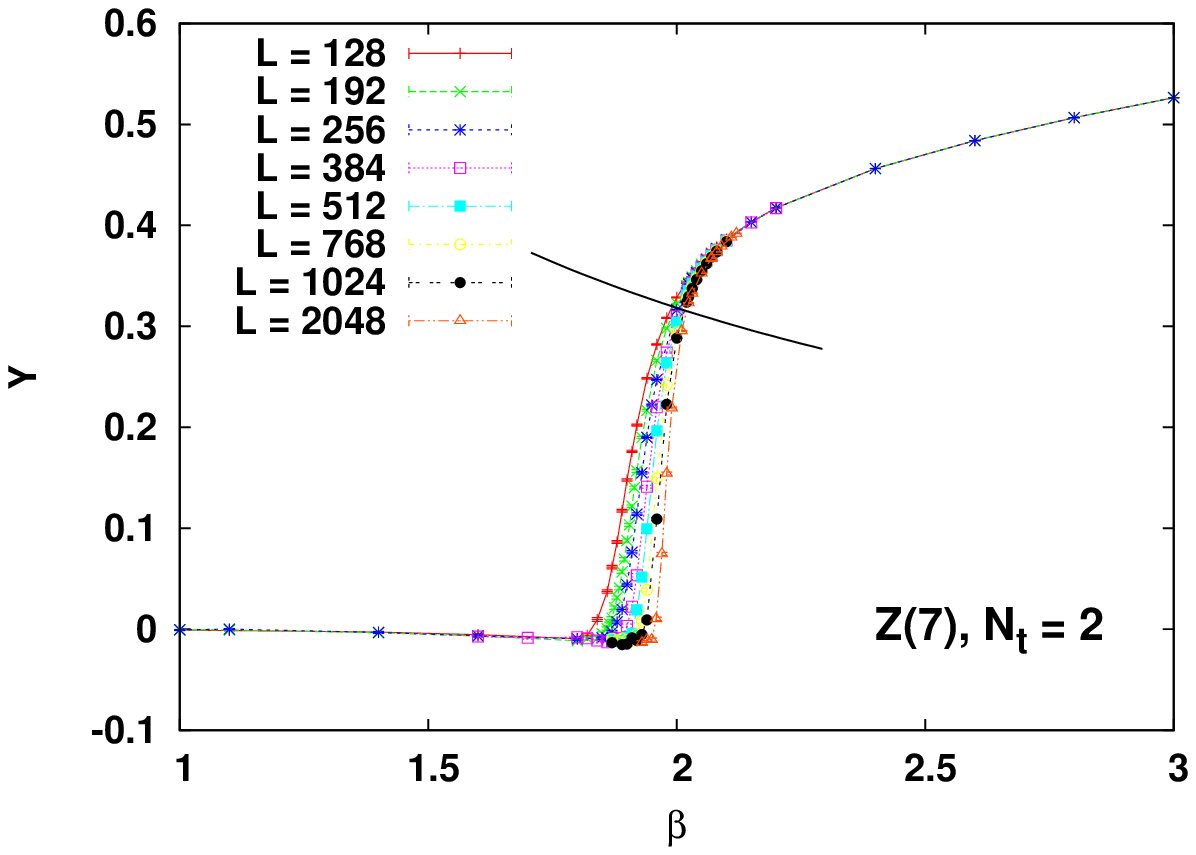}
\end{center}
\caption{Behavior of the helicity modulus $\Upsilon$ for two models considered,
as an illustration of the method (c); the solid line represents the curve
$\Upsilon = 1/(2\pi\beta\eta),\ \eta = 1/4$.}
\label{figura3}
\end{figure}

We show in Figs.~\ref{figura1},~\ref{figura2} and~\ref{figura3} the behavior 
of some of these observables and the method adopted to locate the critical 
couplings. All Tables with the estimations are collected in~\cite{g1}.

\subsection{Determination of the critical indices}

After the determination of the $\beta_{c}$ couplings we can extract 
some critical indices and check the hyperscaling relation 
$d=2\beta / \nu + \gamma / \nu$, where $d$ is the dimension of the system.
For the first transition, according to the standard FSS theory 
the magnetization $|M_{L}|$ at criticality should obey the 
relation $|M_{L}| \sim L^{-\beta / \nu}$ for $L$ large enough. 
Taking into account the possibility of logarithmic corrections we use
\begin{equation}
|M_{L}|=A L^{-\beta/\nu} \ln^r L \ \ \mbox{and} 
\ \ \chi^{(M)}_{L}= A L^{\gamma/\nu} \ln^r L \;,
\label{magn_fss}
\end{equation}
where the second formula is for the susceptibility $\chi^{(M)}_L$, 
with $\gamma/\nu=2-\eta$ ($\eta$ magnetic critical index).

We apply the same procedure to the second transition with the difference
that the fit with the scaling laws Eqs.~(\ref{magn_fss}) is to be applied to 
data for $M_R$ and $\chi_L^{(M_R)}$, respectively.
We found, in general, a reasonable agreement with the expectations (all the 
details and tables can be found in~\cite{g1}).

\begin{figure}
\begin{center}
\includegraphics[scale=0.50]{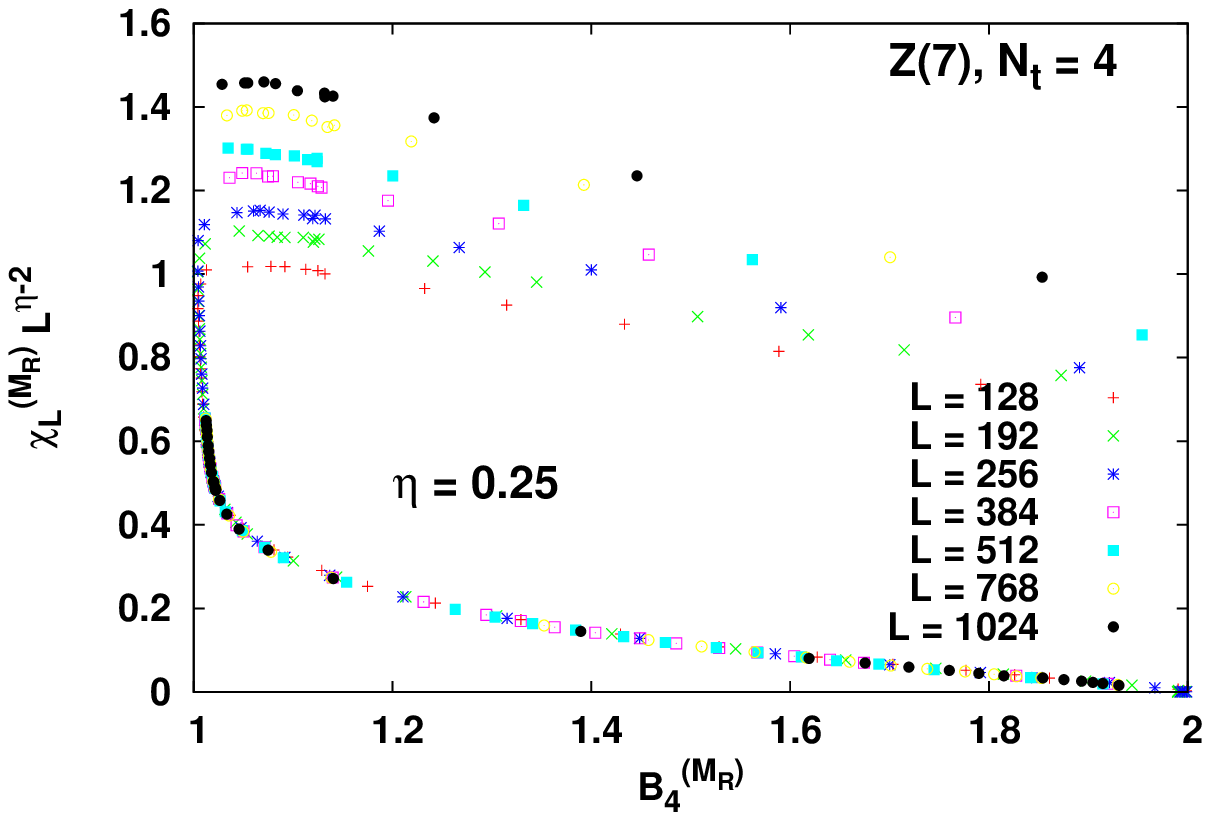}
\includegraphics[scale=0.50]{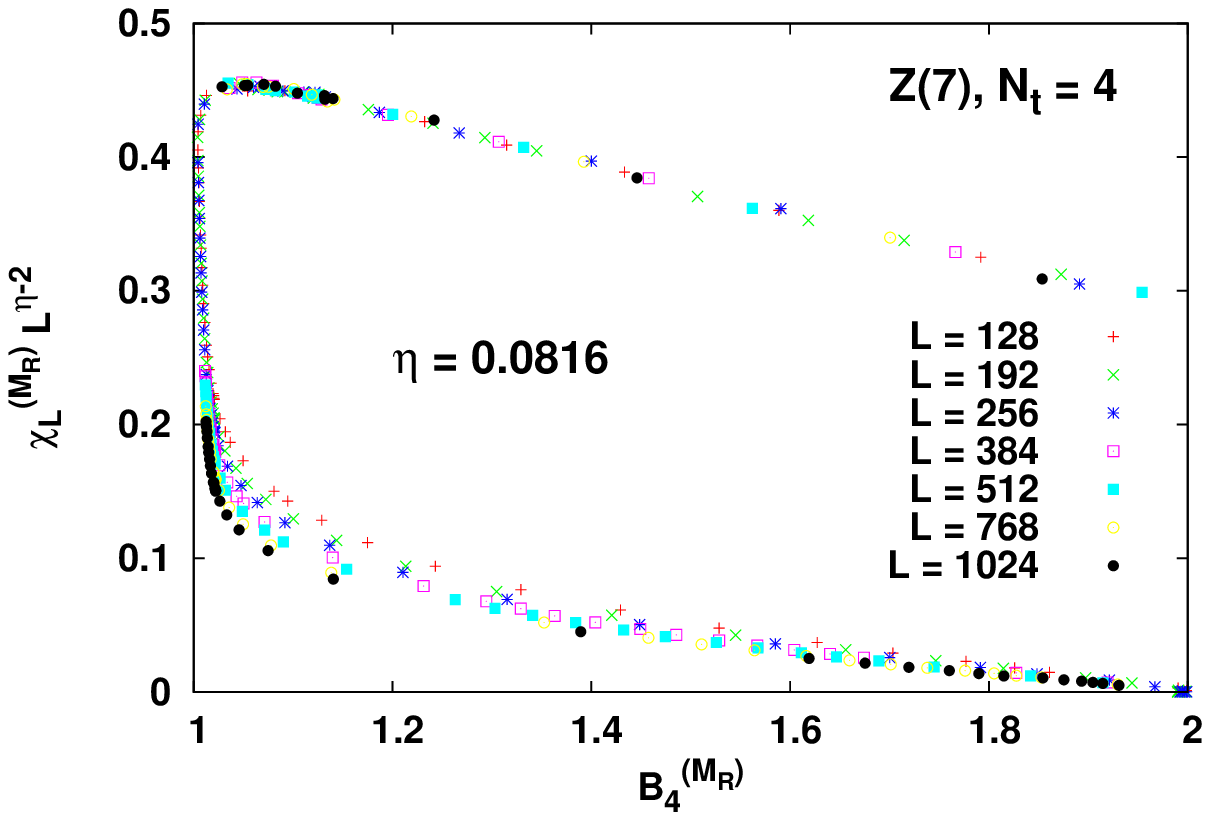}

\includegraphics[scale=0.50]{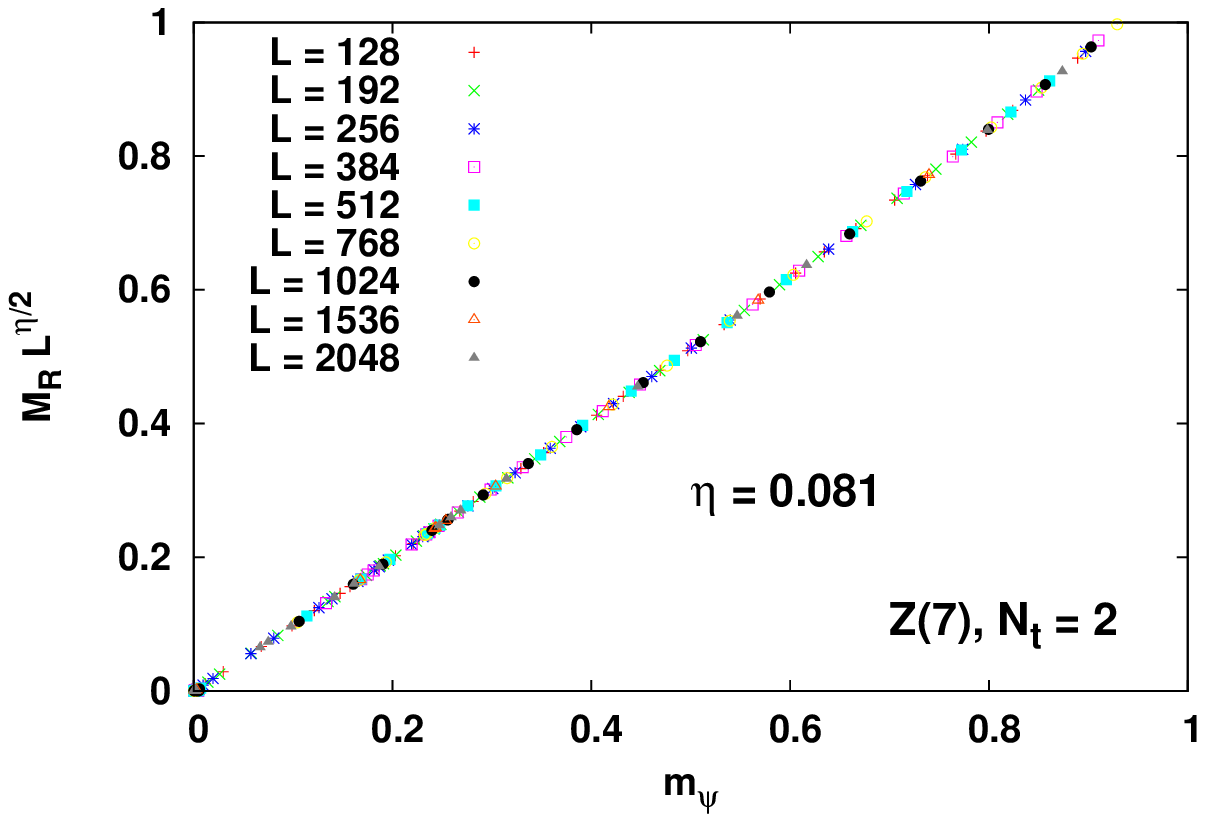}
\includegraphics[scale=0.50]{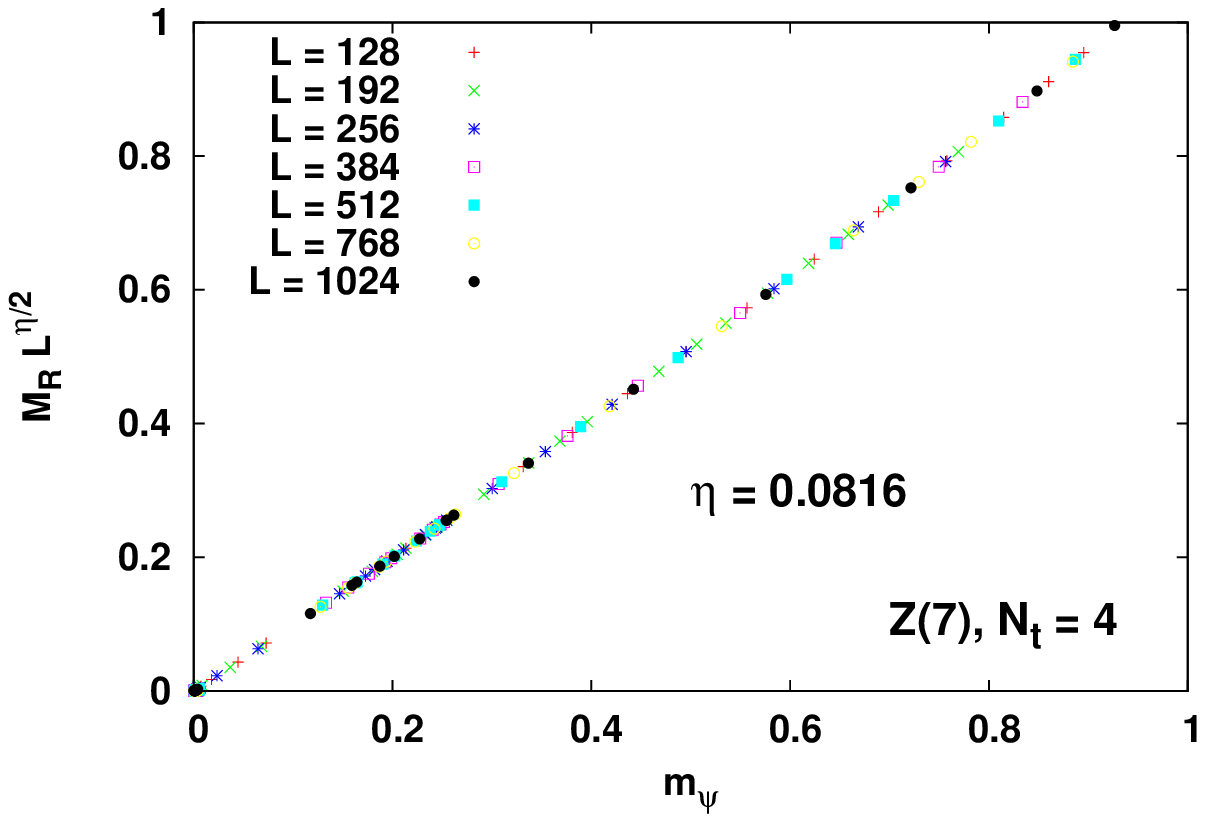}
\end{center}
\caption{Correlation between $\chi_{L}^{(M_{R})}L^{\eta-2}$ and the Binder 
cumulant $B_{4}^{(M_{R})}$ in $Z(7)$ 
with $N_{t}=4$ for $\eta=0.25$ (up and left) and for $\eta=0.0816$ (up and 
right) for different lattice sizes. $M_{R}L^{\eta/2}$ versus 
$m_{\psi}$ in $Z(7)$ with $N_{t}=2$ (down and left) and in $Z(7)$ with 
$N_{t}=4$ (down and right) for $\eta=0.0816$ for different values of lattice 
size $L$.}
\label{figura4}
\end{figure}  

The critical exponent $\eta$, for both transitions in these models
can be calculated without knowledge of the critical temperature, 
building a suitable universal quantity~\cite{Loison}. We show in 
Fig.~\ref{figura4} some results of the behavior of these RG invariant 
quantities for the $Z(7)$ model with $N_{t}=2,4$.
The results for the $\eta$ index is consistent with the FSS method.

\section{Summary and Outlook}

We have determined the two critical couplings of $Z(N=5, 7, 9,13)$ 
LGT and given estimates of the critical indices $\eta$ at both transitions. 
Our findings support for all $N\geqslant 5$ the scenario of three phases: 
a disordered phase at high temperatures, a massless or BKT one at intermediate 
temperatures and an ordered phase, occurring at lower and lower temperatures 
as $N$ increases.
This matches perfectly with the $N\to\infty$ limit, {\it i.e.} the $3D$ 
$U(1)$ LGT (at $\beta_s=0$), where the ordered phase is absent.
We have found that the values of the critical index $\eta$ at the two 
transitions are compatible with the theoretical expectations.
The index $\nu$ also appears to be compatible with the value $1/2$, 
in agreement with RG predictions (see~\cite{g1} and~\cite{g2}).   

Considering the determinations of the critical couplings as a function of $N$, 
we have also conjectured the approximate scaling for $\beta_c^{(1,2)}(N)$. We 
found that $\beta_{\rm c}^{(1)}$ 
converges to the $XY$ value very fast, like $\exp(-a N^2)$ and 
$\beta_{\rm c}^{(2)}$ diverges like $N^2$ (see~\cite{g1}). 

On the basis of these results we are prompted to conclude that 
finite-temperature $3D$ $Z(N)$ LGT for $N>4$ undergoes two phase transitions 
of the BKT type and this model also belongs to the universality class of the 
$2D$ $Z(N)$ spin models, at least in the strong coupling limit $\beta_s=0$.

The new numerical techniques tested in these models can be useful to study 
other models of interest like the $3D$ $SU(N)$ LGT at finite 
temperature and in the strong coupling region.

\acknowledgments
The work of O.B. was supported by the Program of Fundamental Research of the 
Department of Physics and Astronomy of NAS, Ukraine. The work of G.C. and 
M.G. was supported in part by the European Union under ITN STRONGnet (grant 
PITN-GA- 2009-238353).


\end{document}